\documentclass[conference,compsoc]{IEEEtran}

\usepackage[nocompress]{cite}

\usepackage[pdftex]{graphicx}
\graphicspath{{./figure/}}
\DeclareGraphicsExtensions{.pdf,.jpg,.png}

\usepackage[top=10truemm,bottom=10truemm,left=12truemm,right=12truemm]{geometry}
\usepackage{mathptmx} 

\title{PEZY-SC3: A MIMD Many-core Processor for Energy-efficient Computing}

\author{
    \IEEEauthorblockN{
        Naoya Hatta\IEEEauthorrefmark{1},
        Shuntaro Tsunoda\IEEEauthorrefmark{1},
        Kouhei Uchida\IEEEauthorrefmark{1},
        Taichi Ishitani\IEEEauthorrefmark{1},
        Ryota Shioya\IEEEauthorrefmark{1}\IEEEauthorrefmark{2}, and
        Kei Ishii\IEEEauthorrefmark{1}
    }
    \IEEEauthorblockA{
        \IEEEauthorrefmark{1}
        PEZY Computing, K.K. 
        \hspace{1em}
        \IEEEauthorrefmark{2}
        The University of Tokyo 
    }
    \IEEEauthorblockA{
        Email: \{hatta, tsunoda, uchida, ishitani, ishii\}@pezy.co.jp, shioya@ci.i.u-tokyo.ac.jp 
    }
}

\begin{document}

\maketitle

\section{Introduction}

PEZY-SC3 is a highly energy- and area-efficient processor for supercomputers developed using TSMC 7\-nm process technology. 
It is the third generation of the PEZY-SCx series developed by PEZY Computing, K.K.
Supercomputers equipped with the PEZY-SCx series have been deployed at several research centers and are used for large scale scientific calculations~\cite{HOSONO2020, HISHINUMA2020, MATSUMOTO2019, TANAKA2018, IWASAWA2019}.

PEZY-SC3 outperforms previous PEZY-SCx and other processors in terms of energy and area efficiency. 
To achieve high efficiency, PEZY-SC3 employs a MIMD many-core, fine-grained multithreading, and non-coherent cache, focusing on applications involving high thread-level parallelism. 
Our MIMD many-core-based architecture achieves high efficiency while providing higher programmability than existing architectures based on specialized tensor units with limited functionality or wide-SIMD~\cite{Fugaku2020}. 
Another key point of this architecture is to achieve both high efficiency and high throughput without using complex and expensive units such as out-of-order schedulers. 
Moreover, our novel non-coherent and hierarchical cache system enables high scalability on many-core without compromising programmability.

The energy efficiency of a system equipped with PEZY-SC3 is approximately 24.6 GFlops/W as measured by LINPACK, and it ranked 12th in the Green500~\cite{green500} (November 2021), which measures the energy efficiency of supercomputers.
In terms of processor architecture, all the systems ranked higher than the PEZY-SC3 system are equipped with NVIDIA A100 or Preferred Networks MN-Core, and thus PEZY-SC3 is the third-ranked processor after them.
While A100 and MN-Core achieve high energy efficiency with tensor units specialized for specific functions, PEZY-SC3 does not have such specialized tensor units and thus has higher programmability.
Furthermore, the program in systems with PEZY-SC3 was not yet fully optimized, and there is still ample potential for energy efficiency improvements.


\section{Structure}

\begin{figure}[b]
    \centering
    \includegraphics[width=3.8in]{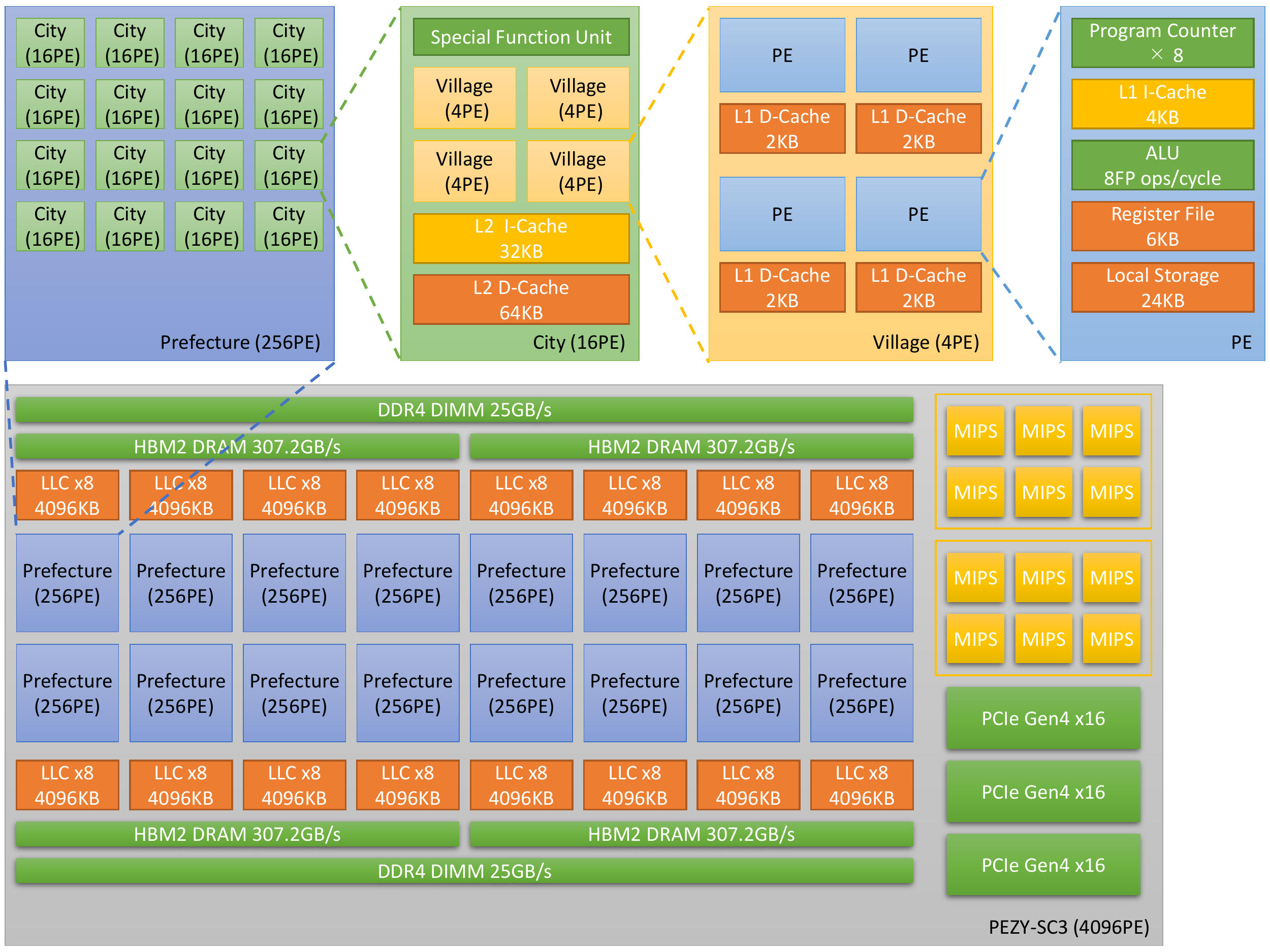}
    \caption{PEZY-SC3 block diagram.}
    \label{block}
\end{figure}

Figure~\ref{block} shows a PEZY-SC3 block diagram, and Table~\ref{spec_comp} shows the specifications of each unit comprising PEZY-SC3. 
PEZY-SC3 is composed of the following units:

\begin{itemize}

\item Processor Element (PE):
    PE is the primary computing resource with a custom RISC-like instruction set architecture that we have developed.
    It supports integer and half/single/double precision floating-point arithmetic operations.
\item Management Processor (MP): 
    MP is a processor with MIPS64 ISA that controls the PEs and PCIe interfaces.
    PEZY-SC3 has two clusters of MPs. 
\item External Memory:
    PEZY-SC3 supports two types of external memory: DDR4 and HBM2.
\item 
    External Interface: PEZY-SC3 supports PCIe Gen4 as an external interface.
\end{itemize}

\begin{table*}[htb]
    \centering
    \caption{PEZY-SC3 specification.}
    \begin{tabular}{|l|l||l|l|l|}
        \hline
        \multicolumn{2}{|l||}{}                   & PEZY-SC3                            & PEZY-SC2 (Previous Version)                    \\
        \hline
        \hline
                             & ISA                & Custom ISA                          & Custom ISA                    \\
        Processor Element    & Number of PEs      & 4096                                & 2048                        \\
                             & Frequency          & 1.2\,GHz                            & 1.0\,GHz                    \\
        \hline
                             & ISA                & MIPS64                              & MIPS64                      \\
        Management Processor & Number of MPs      & 6 $\times$ 2cluster                 & 6 $\times$ 1cluster         \\
                             & Frequency          & 1.5\,GHz                            & 1.0\,GHz                    \\
        \hline
                             & Double Precision   & 19.7\,TFlops                        & 4.1\,TFlops                 \\
        Peak Performance     & Single Precision   & 39.3\,TFlops                        & 8.2\,TFlops                 \\
                             & Half Precision     & 78.6\,TFlops                        & 16.4\,TFlops                \\
        \hline
        \multicolumn{2}{|l||}{External Memory}    & DDR4-3200 2ch (51.2\,GB/s)          & DDR4-3200 4ch (102.4\,GB/s) \\
        \multicolumn{2}{|l||}{}                   & HBM2 2.4\,Gbps 4devices (1.2\,TB/s) &                             \\
        \hline
        \multicolumn{2}{|l||}{External Interface} & PCIe Gen4 48lane (96\,GB/s)         & PCIe Gen4 32lane (64\,GB/s) \\
        \hline
    \end{tabular}
    \label{spec_comp}
\end{table*}

\section{Microarchitecture}

PEZY-SC3 has a hierarchical structure comprising units called \textit{prefectures}, \textit{cities}, and \textit{villages}.
The entire chip consists of 16 sets of prefectures and a 4-MB last-level cache (LLC).
Each prefecture consists of 16 cities. 
Each city consists of four villages, a special function unit, a 32-KB L2 instruction cache, and a 64-KB L2 data cache. 
Each village consists of a PE and a 2-KB L1 data cache.

The PE is a fine-grained multithreading processor with eight program counters.
It has a 4-KB L1 instruction cache and 24-KB local storage.
Each PE can issue up to two instructions in each cycle and has two thread groups, each with four threads. 
The PE activates one thread group and executes all four threads in an activated group simultaneously. 
A programmer explicitly switches the activated group using special instructions. 
Through this thread switching mechanism, the PE can effectively hide long memory latency.

\section{Implementation}

\begin{figure}[t]
    \centering

    \includegraphics[width=2.8in]{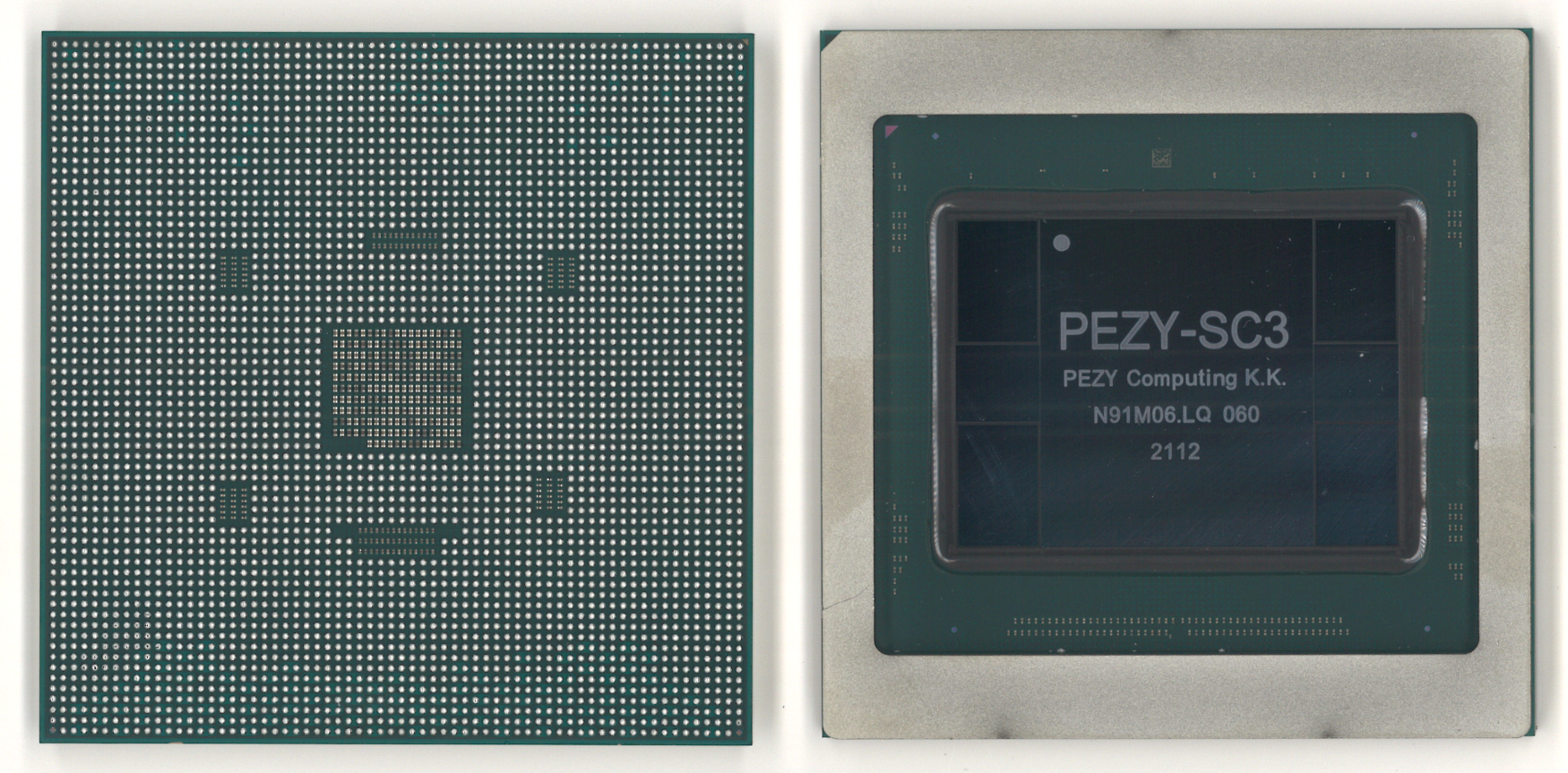}
    \caption{PEZY-SC3 chip package.}
    \label{chip}

    \vspace{0.3in}

    \includegraphics[width=2.8in]{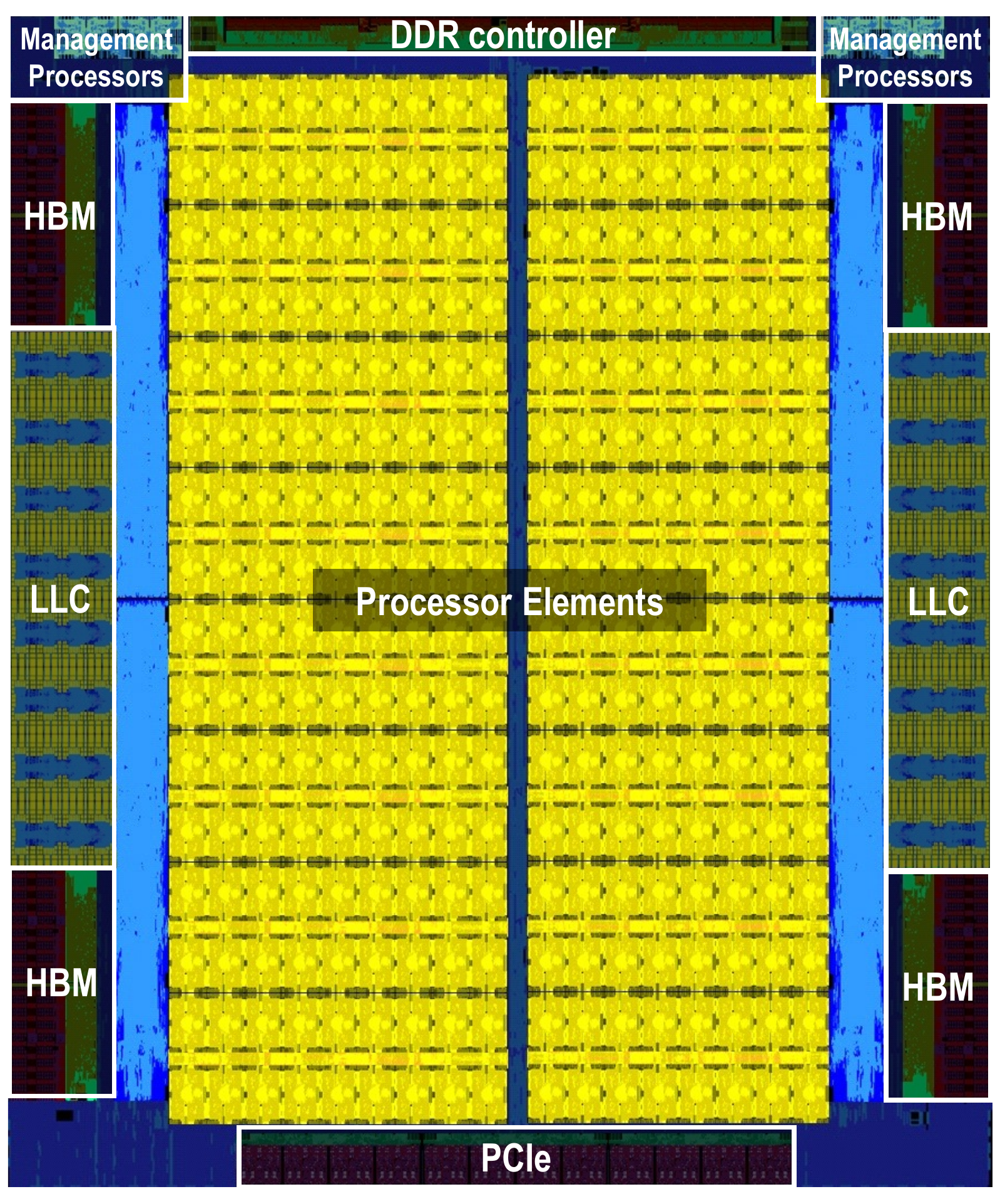}
    \caption{PEZY-SC3 chip layout.}
    \label{layout}
\end{figure}

Table\ref{impl} summarizes the implementation results of PEZY-SC3. 
The TSMC 7-nm process was adopted for PEZY-SC3, and the die size was 25.7\,mm $\times$ 30.6\,mm without scribe lines. 
Figures \ref{layout} and \ref{layout} show a PEZY-SC3 chip and the final GDS of PEZY-SC3. 
The central area of the chip is occupied by the PEs. 
The HBM2 interfaces and LLCs are placed on the left and right edges, respectively. 
DDR interfaces are placed at the center of the top edge, and the MPs are placed on the left and right sides of the top edge. 
Finally, the PCIe interfaces are placed at the bottom edge.

\begin{table}[htb]
    \centering
    \caption{PEZY-SC3 implementation.}
    \begin{tabular}{|l|l|}
        \hline
        Process           & TSMC 7\,nm FinFET          \\
        \hline
        Die Size          & 25.7\,mm $\times$ 30.6\,mm \\
        \hline
        Gate Count        & 3300M gates                \\
        \hline
        Memory Bit Count  & 2300M bits                 \\
        \hline
        Power Consumption & 470\,W (Max)               \\
        \hline
    \end{tabular}
    \label{impl}
    
    \vspace{0.3in}
    
    \caption{System configuration.}
    \begin{tabular}{|l|l|}
        \hline    Number of Nodes       & 50 nodes \\
        \hline    Host Processor        & AMD EPYC 7702P $\times$ 1 for each node \\
        \hline    Processor             & PEZY-SC3 $\times$ 4 for each node\\
        \hline    Total Number of PEs   & 819,200 PEs \\
        \hline    Interconnect          & EDR Infiniband \\
        \hline    Rmax (TFlops/s)       &  1,684.83   \\
        \hline    Rpeak (TFlops/s)      &  2,353.85   \\
        \hline
    \end{tabular}
    \label{system}

\end{table}

\section{Performance and Energy Efficiency}

The measured power consumption for calculating the matrix multiplication with double precision was 300.4\,W when the operating frequency is 800MHz. The chip energy efficiency is 28.45\,GFlops/W.

We also measured the performance and energy efficiency of a system equipped with PEZY-SC3 by LINPACK according to the Top500 regulations. 
The system configuration we used is summarized in Table~\ref{system}. 
The effective performance of our system (Rmax) is 1,684.83 TFlops/s while the peak performance (Rpeak) is 2,353.85 TFlops/s.
The energy efficiency of our system was about 24.6 GFlops/W, and it ranked 12th in the Green500~\cite{green500} (November 2021).

\section{Conclusion}

PEZY-SC3 is a MIMD many-core processor designed for energy-efficient supercomputers and developed using TSMC 7\-nm process technology.
It achieved high energy efficiency while having high programmability.
The energy efficiency of the system equipped with PEZY-SC3 is approximately 24.6 GFlops/W.


\bibliographystyle{IEEEtran}
\bibliography{IEEEabrv,main}

\end{document}